\documentclass[%
 aps,
 prl,
amsmath,amssymb,
reprint,
]{revtex4-1}

\usepackage{graphicx}
\usepackage{dcolumn}
\usepackage{bm}

\usepackage[utf8]{inputenc}
\usepackage{mathptmx}
\usepackage{multirow}
\usepackage{braket}

\usepackage[english]{babel}
\usepackage{xcolor}
\colorlet{RED}{red}
\colorlet{BLUE}{blue}
\usepackage{dcolumn}
\usepackage{bm}
\usepackage[version=4]{mhchem} 
\usepackage{acronym}
\usepackage{adjustbox}
\usepackage{float}
\usepackage{tikz}
\usepackage{microtype} 
\usepackage{algorithm}
\usepackage{xcolor}
\usepackage{algpseudocode}
\usepackage{soul}

\usetikzlibrary{calc,shapes.geometric,decorations.pathmorphing,patterns}

\definecolor{background-color}{gray}{0.98}
\usepackage[margin=2.3cm,bmargin=1cm,footnotesep=1cm]{geometry}

\begin{document}

\title{
Resource-adaptive quantum flow algorithms for quantum simulations of many-body systems: sub-flow embedding procedures
}

\author{Karol Kowalski}
\email{karol.kowalski@pnnl.gov}
\affiliation{%
  Physical Sciences Division, 
  Pacific Northwest National Laboratory, Richland, Washington, 99354, USA
}

\author{Nicholas P. Bauman}
\affiliation{%
  Physical Sciences Division, 
  Pacific Northwest National Laboratory, Richland, Washington, 99354, USA
}

\begin{abstract}


In this study, we utilized the quantum flow (QFlow) method to perform quantum simulations of correlated systems. The QFlow approach allows for sampling large sub-spaces of the Hilbert space by solving coupled variational problems in reduced dimensionality active spaces. Our research demonstrates that the circuits for evaluating the low dimensionality subproblems of the QFlow algorithms on quantum computers are significantly less complex than the parent (large subspace of the Hilbert space) problem, opening up possibilities for scalable and constant-circuit-depth quantum computing. Our simulations indicate that QFlow can be used to optimize a large number of wave function parameters without an increase in the required number of qubits. We were able to showcase that a variation of the QFlow procedure can optimize 1,100 wave function parameters using modest quantum resources. Furthermore, we investigated an adaptive approach known as the sub-flow approach, which involves a limited number of active spaces in the quantum flow process. Our findings shed light on the potential of QFlow in efficiently handling correlated systems via quantum simulations.
\end{abstract}

\maketitle

\section{Introduction}

Many areas of computational chemistry and physics demand design principles that can effectively reduce the dimensionality of the quantum problems to make them numerically tractable in conventional or quantum simulations. Over the last decade, one could witness significant progress in enabling various techniques, especially in the context of developing embedding schemes, \cite{RevModPhys.78.865_2006,PhysRevLett.109.186404.2012,lan2017generalized,lin2022variational}  methods generally attributed to the mean-field approaches, downfolding algorithms \cite{10.1063/9780735422490_2022,romanova2023dynamical,chang2024downfolding}. A common feature of all these approaches is the careful treatment of correlation effects, which define various sub-systems and interactions between them. 
An excellent illustration of the progress achieved can be provided by various tensor-product state-based approaches such as the block-correlated coupled cluster approach \cite{li2004block}, the Cluster Mean-Field theory \cite{jimenez2015cluster,papastathopoulos2023symmetry}, the Active Space Decomposition method \cite{parker2013communication,parker2014communication}, the Variational Localized Active Space Self Consistent field-State Interaction method \cite{hermes2019multiconfigurational,hermes2020variational}, and the Tensor Product-state Selected CI and Tensor Product State-Coupled Electron Pair Approximation approaches \cite{abraham2020selected,abraham2022coupled}.

Recently, new properties of single-reference coupled cluster (SR-CC)  method \cite{kowalski2018properties,kowalski2021dimensionality,kowalski2023sub} associated with the existence of the so-called sub-system embedding sub-algebras (SES) and the possibility of calculating CC energies as eigenvalues of the effective Hamiltonians in the SES-generated active spaces (we will refer to as the CC downfolding) have provided a novel design principle for embedding and renormalization procedures. The SR-CC downfolding also provides a natural language for multi-site embedding known as Equivalence Theorem discussed in Refs.\cite{kowalski2018properties,kowalski2021dimensionality}.
In the last few years, the idea of CC downfolding was exploited to construct embedding schemes, including Second-Order Active-Space Embedding Theory (ASET(2)) \cite{he2022second}  and many-body perturbation coupled-cluster (MB-CC) a static quantum embedding scheme \cite{shee2024static}.

The possibility of forming accurate downfolded/effective Hamiltonians or reducing the dimensionality of quantum problems has also drawn much attention in quantum computing. The primary motivation for this effort is the possibility of utilizing NISQ-era technology to perform quantum simulations for more realistic systems/processes than when using available registers of qubits for bare Hamiltonians. 
One of the promising ways for developing scalable and constant circuit-depth quantum algorithms is the recently introduced hybrid Quantum Flow (QFlow) algorithm \cite{kowalski2021dimensionality,kowalski2023quantum} (see also Ref.\cite{kowalski2018properties} for early formulations of CC flow methods). The QFlow approach utilizes classical and quantum resources to solve coupled variational problems in small-dimensionality active spaces. While the classical computer is used as a storage device and generator of effective Hamiltonians for active spaces involved in the flow, the quantum computer(s) is(are) used to optimize various active-space problems using quantum algorithms, including ubiquitous variational quantum eigensolver (VQE) algorithm \cite{peruzzo2014variational,mcclean2016theory,romero2018strategies,PhysRevA.95.020501,Kandala2017,kandala2018extending,PhysRevX.8.011021,huggins2020non,cao2019quantum}.
Among several advantages of the QFlow algorithm is the possibility of sampling large sub-spaces of Hilbert space using modest quantum resources. The QFlow algorithm is also well suited to build parallel/distributed hybrid algorithms and to capture various types of sparsity in quantum systems. Previous studies demonstrated that for the H8 model in the STO-3G basis set, one can optimize 684 parameters using procedures that require eight qubits. In this paper, we will provide a further extension of the QFLow procedure by (1) illustrating the application of the QFlow algorithm to problems that require the optimization of  1,100 parameters and (2) introducing the sub-flow (sub-QFlow) based algorithms that utilize the limited number of active-space problems. The former approach provides an adaptive framework to include in QFlow in the case when the inclusion of all possible active spaces of a given size is too expensive. We also outline the integration of the QFlow and sub-QFlow approaches with perturbative expansion. This analysis is motivated by the general features of the Bloch equation for the wave operator formalism.
%
%
%
%

\section{Coupled Cluster Downfolding Theory and Bloch equation formalism}

An alternative way of introducing the CC downfolding goes through the analysis of the wave-operator-based Bloch equation \cite{bloch1958theorie,jorgensen1975effective,durand1983direct,jeziorski1981coupled,lindgren1987connectivity,piecuch1994application,meissner1995effective,meissner1998fock,kowalski2000complete,pittner2003continuous,li2003general,lindgren2012atomic,shavitt2009many}, which also provides a theoretical platform for theoretical analysis of perturbative expansions and properties of various parametrizations of the wave operator. 
Therefore, our goal is to integrate the many-body perturbation theory with recently introduced ideas of quantum flows in a way that the perturbative arguments can be integrated with the iterative framework defined by various active-space problems in a way that assures the accuracy and numerical stability of the proposed algorithm. Although, as mentioned earlier, there are numerous approaches of this type, the proposed algorithm follows different design principles based on the recently introduced CC downfolding techniques and the possibility of calculating CC energies in an alternative way compared to the standard CC energy expression. The CC downfolding has been discussed both for the standard  CC theory and general unitary CC (GUCC) approaches. While the former approach uses commuting operators to define the corresponding cluster operator, the latter case is more challenging due to the presence of a non-commuting operator in the anti-Hermitian cluster operator. 
The possibility of finding alternative ways of expressing energies and integrating perturbative and iterative formulations can be illustrated in the example of specific solutions of the Bloch equations often used to derive linked cluster theorem employing wave operator formalism. Following Adamowicz, Oliphant, and Piecuch's ideas 
\cite{oliphant1991multireference,oliphant1992implementation,pnl93,piecuch1994state}
we assume that the general wave operator $\Omega$  for a single-reference case  can be factorized as:
\begin{equation}
\Omega |\Phi\rangle = \Omega_{\rm ext} \Omega_{\rm int} |\Phi\rangle \;,
\label{eq1}
\end{equation}
where $|\Phi\rangle$ is a reference Slater determinant, $\Omega$ is the wave operator,
$\Omega_{\rm int}$ is an operator capable of producing full configuration interaction type  expansion within the pre-defined active space when acting onto $|\Phi\rangle$,
$\Omega_{\rm ext}$ is operator producing all possible excitation in the orthogonal component of the active space when acting on $|\Phi\rangle$. We assume that the $\Omega$ operator satisfies the  intermediate normalization condition, and the inverse of their second quantized form of $\Omega_{\rm ext}$ can be defined in the Fock space.
The $P$, $Q_{\rm int}$, and  $Q_{\rm ext}$ operators designate the projection operators onto reference function ($P=|\Phi\rangle\langle\Phi|$), the space of all excited Slater determinants in the active space, and the space of all excited Slater determinants from the orthogonal complement of the active space such that 
\begin{equation}
    P+Q = P+Q_{\rm int}+Q_{\rm ext}= 1 \;,
    \label{eq2}
\end{equation}
where $1$ stands for the identity operator in the Hilbert space. Therefore, $P+Q_{\rm int}$ operator gives the projection onto the active space of interest.  Now, we insert wave operator (\ref{eq1}) into the Bloch equation
\begin{equation}
    [\Omega,H_0]|\Phi\rangle = V\Omega |\Phi\rangle - \Omega P V \Omega |\Phi\rangle
\label{bloch}
\end{equation}
where the Hamiltonian $H$ is a sum of 0-th order Hamiltonian ($H_0$) and perturbation $V$, i.e., $H=H_0 + V$. We also assume that the reference function is an eigenstate of $H_0$, $H_0|\Phi\rangle = E_0 |\Phi\rangle$, where $E_0$ is 0-th order energy. Assuming that the inverse of the  $\Omega_{\rm ext}$ operator
($\Omega_{\rm ext}^{-1}$) exist, after  projecting the Bloch equation onto active space ($P+Q_{\rm int}$) and its orthogonal complement ($Q_{\rm ext}$), one arrives at the equivalent form of the Bloch equations:
\begin{widetext}
\begin{eqnarray}
    (P+Q_{\rm int}) \Omega_{\rm ext}^{-1}H\Omega_{\rm ext}(P+Q_{\rm ext}) \Omega_{\rm int}|\Phi\rangle &=& E \Omega_{\rm int}|\Phi\rangle \;,
    \label{blochd1} \\
    Q_{\rm ext}[\Omega_{\rm ext}\Omega_{\rm int},H_0]|\Phi\rangle &=& 
    Q_{\rm ext}V\Omega_{\rm ext}\Omega_{\rm int}|\Phi\rangle - Q_{\rm ext}\Omega_{\rm ext}\Omega_{\rm int}|\Phi\rangle V \Omega_{\rm ext}\Omega_{\rm int}|\Phi\rangle|\Phi\rangle \;,\label{blochd2}
\end{eqnarray}
\end{widetext}
where $E$ is the exact energy of the quantum system and 
\begin{equation}
(P+Q_{\rm int})\Omega_{\rm int}|\Phi\rangle=\Omega_{\rm int}|\Phi\rangle \;.
\label{aux1}
\end{equation}
The representation of the Bloch equations for wave operator representation given by  Eq.(\ref{eq1}) provides a downfolded form of the effective Hamiltonian $H^{\rm eff}(D)$
\begin{equation}
H^{\rm eff}(D) = (P+Q_{\rm int}) \Omega_{\rm ext}^{-1}H\Omega_{\rm ext}(P+Q_{\rm ext})\;. \label{aux3}
\end{equation}
It should be compared to the standard form of the effective Hamiltonian ($H^{\rm eff}(B)$) defined in the Bloch formalism as
\begin{equation}
H^{\rm eff}(B) = PH\Omega_{\rm ext}\Omega_{\rm int}P \;, \label{aux2}
\end{equation}
which for $\Omega=\Omega_{\rm ext}\Omega_{\rm int}$ being a solution to Bloch equation defines energy $E$ of the system.
Using this definition, the Bloch equation (\ref{blochd1}-\ref{blochd2}) can be written as
\begin{widetext}
\begin{eqnarray}
    H^{\rm eff}(D)\Omega_{\rm int}|\Phi\rangle &=& E \Omega_{\rm int}|\Phi\rangle \;,
    \label{blochd11} \\
    Q_{\rm ext}[\Omega_{\rm ext}\Omega_{\rm int},H_0]|\Phi\rangle &=& 
    Q_{\rm ext}V\Omega_{\rm ext}\Omega_{\rm int}|\Phi\rangle - Q_{\rm ext}\Omega_{\rm ext}\Omega_{\rm int}|\Phi\rangle V \Omega_{\rm ext}\Omega_{\rm int}|\Phi\rangle|\Phi\rangle \;.\label{blochd22}
\end{eqnarray}
\end{widetext}
The alternative form of Bloch equations for a class of factorized wave operators  (\ref{eq1}) also suggests alternative ways of solving Bloch equations combining perturbative solvers for $\Omega_{\rm ext}$, Eq.(\ref{blochd22}), and iterative approaches to solve eigenvalue problem given by Eq. (\ref{blochd11}).
We will explore these hybrid formulations, which correspond to the embedding of variational solvers into the perturbative formulations in the context of quantum sub-flow formulations.

 A good example illustrating factorization (\ref{eq1})  for particular classes of active spaces is provided by the  single-reference CC Ansatz
 \begin{equation}
     \Omega|\Phi\rangle = e^T |\Phi\rangle \;,
     \label{aux5}
 \end{equation}
 where $T$ is the cluster operator. In Ref.\cite{kowalski2018properties} we demonstrated that sub-system embedding sub-algebras (SES; generally designated as $\mathfrak{h}$) induce the partitioning of the cluster operator into commuting  internal $T_{\rm int}(\mathfrak{h})$ (belonging to $\mathfrak{h}$)  and external $T_{\rm ext}(\mathfrak{h})$ parts, i.e.,
 \begin{equation}
     T = T_{\rm int}(\mathfrak{h})+T_{\rm ext}(\mathfrak{h})
     \label{part1}
 \end{equation}
 such that $\Omega_{\rm ext}$ and $\Omega_{\rm int}$ can be expressed as 
 \begin{eqnarray}
     \Omega_{\rm ext} &=& e^{T_{\rm ext}(\mathfrak{h})} \;, \label{aux6} \\
     \Omega_{\rm int} &=& e^{T_{\rm int}(\mathfrak{h})} \;.
     \label{aux7}
 \end{eqnarray}
The partitioning (\ref{part1}) originates in the active-space CC formulations \cite{pnl93}  and was used by Piecuch and Adamowicz to select higher-order amplitudes for the active-space CC formulations such as CCSDt and CCSDtq models \cite{oliphant1991multireference,oliphant1992implementation,piecuch1999coupled,ivanov2009multireference,piecuch2010active}.  
A similar decomposition of the cluster operator has been  used in the context of the tailored CC formalism 
\cite{kinoshita2005coupled,hino2006tailored}.
The equation (\ref{blochd11}) indicates that the CC energy can be calculated alternatively as an eigenvalue of CC downfolded Hamiltonian $H^{\rm eff}(\mathfrak{h})$ defined as 
\begin{equation}
    H^{\rm eff}(\mathfrak{h}) e^{T_{\rm int}(\mathfrak{h})}|\Phi\rangle = E e^{T_{\rm int}(\mathfrak{h})}|\Phi\rangle \;, \label{aux8}
\end{equation}
where 
\begin{equation}
 H^{\rm eff}(\mathfrak{h}) = (P+Q_{\rm int}(\mathfrak{h}))
 e^{-T_{\rm ext}(\mathfrak{h})} H e^{T_{\rm ext}(\mathfrak{h})}
 (P+Q_{\rm int}(\mathfrak{h})) \;.
 \label{aux9}
\end{equation}
In the equation above, the $Q_{\rm int}(\mathfrak{h})$ is the projection operator onto all excited Slater determinants spanning active space generated by SES $\mathfrak{h}$. The above result (Eq.(\ref{aux8})) is known under SES-Theorem and holds for any SES $\mathfrak{h}$ specific to a given cluster operator $T$.

Based on our earlier  discussion, one can envision an alternative way of solving Bloch equations using CC Ansatz. For example, $T_{\rm ext}(\mathfrak{h})$ can be determined using its low-order perturbative estimates (we will denote $i$-th order contribution to
$T_{\rm ext}(\mathfrak{h})$ as
$T_{\rm ext}^{(i)}(\mathfrak{h})$ ), i.e.,
\begin{equation}
    T_{\rm ext}^{[n]}(\mathfrak{h}) = \sum_{i=1}^{n}  T_{\rm ext}^{(i)}(\mathfrak{h})\;,
    \label{aux10}
\end{equation}
whereas $T_{\rm int}(\mathfrak{h})$ is evaluated iteratively by solving Eq.(\ref{aux8}) 
with perturbative approximation $H^{{\rm eff}[n]}(\mathfrak{h})$ of the effective Hamiltonian $H^{\rm eff}(\mathfrak{h})$ given by the expression:
\begin{equation}
 H^{{\rm eff}[n]}(\mathfrak{h}) = (P+Q_{\rm int}(\mathfrak{h}))
 e^{-T_{\rm ext}^{[n]}(\mathfrak{h})} H e^{T_{\rm ext}^{[n]}(\mathfrak{h})}
 (P+Q_{\rm int}(\mathfrak{h})) \;.
 \label{aux11}
\end{equation}
For well-defined active spaces, the perturbative expansion for  $T_{\rm ext}(\mathfrak{h})$, due to larger energy denominators, should be more numerically stable compared to the perturbative estimates of $T_{\rm int}(\mathfrak{h})$. 
This feature  is a general  advantage of using procedures based on the double  similarity transformations
 \cite{meissner1995effective,meissner1998fock}.


\section{Quantum Flow Formalism}

The quantum flow (QFLow) algorithm  originates in the  
invariance of the CC energy with respect to  the choice of SES and the so-called Equivalence Theorem \cite{kowalski2021dimensionality,bauman2022coupled}, which states that when several SES problems 
represented by (\ref{aux8}) are coupled into the flow, i.e.,
\begin{equation}
H^{\rm eff}(\mathfrak{h}_i) e^{T_{\rm int}(\mathfrak{h}_i)}|\Phi\rangle =  E e^{T_{\rm int}(\mathfrak{h}_i)}|\Phi\rangle  \;, (i=1,\ldots,M) \label{flow1}
\end{equation}
($M$ stands for the number of complete active spaces (CASs) included in the flow), then the corresponding solution of (\ref{flow1}) is equivalent to the solution of the standard representation of the CC theory 
\begin{eqnarray}
Q_{\rm QFlow}e^{-T_{\rm QFlow}}H e^{T_{\rm QFlow}} |\Phi\rangle &=& 0 \;,\label{cceq1} \\
\langle\Phi|^{-T_{\rm QFlow}}H e^{T_{\rm QFlow}} |\Phi\rangle &=& E \;,
\label{cceq2}
\end{eqnarray}
with the $T_{\rm QFlow}$ operator is defined as a 
combination of all 
non-repetitive
excitations included in
$T_{\rm int}(\mathfrak{h}_i)\; (i = 1, . . . , M)$ operators, 
symbolically denoted as,
%
\begin{equation}
  T_{\rm QFlow}= 
  {\widetilde{\sum}}_{i=1}^M 
  T_{\rm int}(\mathfrak{h}_i) \;.
  \label{flow2}
\end{equation}
and the external cluster operator with respect to the $\mathfrak{h}_i$-generated CAS is defined as
\begin{equation}
  T_{\rm ext}(\mathfrak{h}_i) = T_{\rm QFlow} - T_{\rm int}(\mathfrak{h}_i) \;, (i=1,\ldots,M).
  \label{extTflow}
\end{equation}
In the same way, the $Q_{\rm QFlow}$ operator is the projection operator onto all excited (with respect to $|\Phi\rangle$) Slater determinants generated by $T_{\rm QFlow}$ when acting on the reference function.
It should be stressed that all effective Hamiltonians in (\ref{flow1}) are non-Hermitian.

A Hermitian version of the above form of QFlow algorithm (i.e., all effective Hamiltonians involved in the flow are Hermitian operators)  can be derived by analyzing energy functional defined by general-type anti-Hermitian cluster operator $\sigma$:
\begin{equation}
E = \langle\Phi| e^{-\sigma_{\rm QFlow}} H e^{\sigma_{\rm QFlow}} |\Phi\rangle
\label{var1}
\end{equation}
where one assumes that the $\sigma$ operator for the ground state of interest can be represented as a superposition of anti-Hermitian  cluster operators
$\sigma_{\rm int}(\mathfrak{h}_i)$ defined for 
$\mathfrak{h}_i$-generated CASs, i.e.,
\begin{equation}
  \sigma_{\rm QFlow} \simeq 
  {\widetilde{\sum}}_{i=1}^M 
  \sigma_{\rm int}(\mathfrak{h}_i) \;,
  \label{var2}
\end{equation}
as in Eq.(\ref{flow2}). Additionally, we will assume that each $\sigma_{\rm int}(\mathfrak{h}_i)$ is represented by the unitary CC (UCC) Ansatz, i.e.,
\begin{equation}
\sigma_{\rm int}(\mathfrak{h}_i) \simeq T_{\rm int}(\mathfrak{h}_i) -
T_{\rm int}(\mathfrak{h}_i)^{\dagger} \;.\label{var2a}
\end{equation}
As shown in Ref.\cite{kowalski2023quantum}, using rank-$N$ Trotter formula for each active-space  decompositions 
\begin{equation}
 \sigma_{\rm QFlow} =  \sigma_{\rm int}(\mathfrak{h}_i)+ \sigma_{\rm ext}(\mathfrak{h}_i) \;,
 \label{var3}
\end{equation}
i.e.,
\begin{equation}
e^{\sigma_{\rm QFlow}} \simeq ( e^{\sigma_{\rm ext}(\mathfrak{h}_i)/N}
     e^{\sigma_{\rm int}(\mathfrak{h}_i)/N})^{N} \;,
     \label{var3a}
\end{equation}
we can re-cast optimization problem (\ref{var1}) into the form similar to (\ref{flow1}), representing coupled minimization problems. For example, for the rank-1 ($N=1$) Trotter decomposition, one gets the following form of the Hermitian quantum flow:
\begin{equation}
\min_{\sigma_{\rm int}(\mathfrak{h}_i)} \langle\Phi| e^{-\sigma_{\rm int}(\mathfrak{h}_i)}H^{\rm eff}(\mathfrak{h}_i) e^{\sigma_{\rm int}(\mathfrak{h}_i)}|\Phi\rangle  \;, (i=1,\ldots,M) \label{var4} 
\end{equation}
where each problem is optimized with respect to amplitudes defining the $\sigma_{\rm int}(\mathfrak{h}_i)$ operator.
For rank-1 Trotter decomposition of $e^{\sigma}$/$e^{-\sigma}$ with respect to the partitioning (\ref{var3}), the 
effective Hamiltonians $H^{\rm eff}(\mathfrak{h}_i)$ are defined as
\begin{equation}
H^{\rm eff}(\mathfrak{h}_i)=(P+Q_{\rm int}(\mathfrak{h}_i)) e^{-\sigma_{\rm ext}(\mathfrak{h}_i)} H  e^{\sigma_{\rm ext}(\mathfrak{h}_i)} (P+Q_{\rm int}(\mathfrak{h}_i))  \label{var5}
\end{equation}
The flow equations can be solved using the importance ordering of the active spaces (defined with respect to the strength of encapsulated ground-state correlation effects) and optimization of the unique amplitudes, which, due to the possibility of overlapping active spaces, have not been optimized in preceding active-space problems. As discussed in Refs.\cite{kowalski2021dimensionality,bauman2022coupled,kowalski2023quantum}, the energy of QFlow is recorded for the main active space (or the active space no. 1 in the hierarchy of active spaces), which means that the remaining active-space problems can be viewed as "learning agents" to learn correlation effects outside of the main active space.
Since equations for each active space are expressed in terms of connected diagrams (it can be demonstrated using commutator expansion),  the QFlow energies are size-consistent.

If the exact form of the effective Hamiltonians $H^{\rm eff}(\mathfrak{h}_i)$ can be constructed, then the  Hermitian QFlow is composed of coupled variational problems, and the resulting energy is still variational. The finite-rank commutator many-body expansions for $H^{\rm eff}(\mathfrak{h}_i)$ may, in general, violate the variational character of the QFlow energy. 

\section{The quantum sub-flow expansion}

In the present studies, we will integrate QFlow procedures into the
two-step DUCC-QFlow  \cite{kowalski2023quantum}, where we first build the  CC downfolded Hamiltonians for the target active space procedure based on the o double unitary CC Ansatz (DUCC) \cite{bauman2019downfolding,downfolding2020t} to reduce the size of the problem to the target active space (which in real applications should be beyond the reach of currently existing quantum computing) and then solve the downfolded target active space Hamiltonian using QFlow-type procedures. 

\begin{figure*}
\includegraphics[scale=0.47]{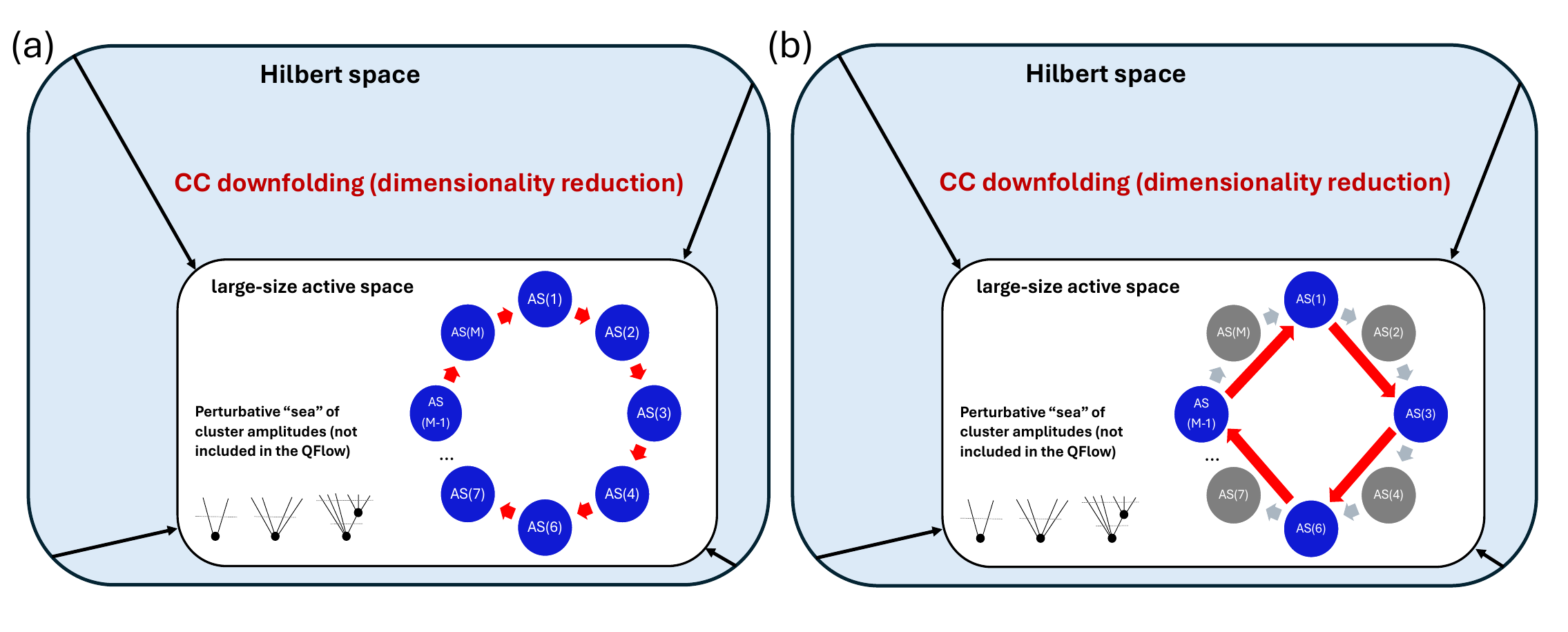}
\caption{
Schematic representation of the quantum sub-flows formalism (see text for details). 
}
\label{fig:subf}
\end{figure*}

In Ref.\cite{kowalski2023quantum}, we explored the QFLow algorithm based on the inclusion of all active spaces of a given type. 
In this paper, as in previous studies, we will employ  $(4e,4o)$ active spaces, 
which is commensurate with the size of typical VQE applications. However, in practical applications, the total number of all possible active spaces of a given type grows rapidly with the size of the target space. Therefore, QFlow approaches, including all active spaces of a given type beyond a certain size of target active space, are not numerically feasible. Instead of considering the set $\mathcal{S}_{\rm QFlow}$ of all fixed size active spaces in the QFlow:
\begin{equation}
    \mathcal{S}_{\rm QFlow} = \lbrace \mathcal{M}_i \rbrace_{i=1}^{M} \;,
    \label{app1}
\end{equation}
one can consider a sub-set of active spaces, $\mathcal{S}_{\rm sub-QFlow}$ defined as
\begin{equation}
   \mathcal{S}_{\rm sub-QFlow} = \lbrace \mathcal{M}_{m_i} \rbrace_{i=1}^{K} \;,
    \label{app2}
\end{equation}
where $K\ll M$ and active spaces in $\mathcal{S}_{\rm sub-QFlow}$ contain the most essential correlation effects. The active spaces of a given type not included in  $\mathcal{S}_{\rm sub-QFlow}$ contain less important correlation effects. 
%
%
%
Here, we will consider two types of anti-Hermitian cluster operators that "embed"  iteratively-determined amplitudes corresponding 
to $\mathcal{S}_{\rm QFlow}$ and $\mathcal{S}_{\rm QFlow}$ activ e spaces into the "sea" of perturbatively determined amplitudes in the target active space, i.e., 
\begin{equation}
  \sigma_{\rm QFlow} \simeq 
  {\widetilde{\sum}}_{i=1}^M 
  \sigma_{\rm int}(\mathfrak{h}_i) + \sigma_M(R)\;,
  \label{expa1}
\end{equation}
and 
\begin{equation}
  \sigma_{\rm sub-QFlow} \simeq 
  {\widetilde{\sum}}_{i=1}^K 
  \sigma_{\rm int}(\mathfrak{h}_i)+
  {\widetilde{\widetilde{\sum}}}_{i=K+1}^M 
  \sigma_{\rm int}(\mathfrak{h}_i) +   
  \sigma_M(R)\;,
  \label{expa2}
\end{equation}
where ${\widetilde{\widetilde{\sum}}}_{i=K+1}^M \sigma_{\rm int}(\mathfrak{h}_i)$ is defined by corresponding excitations/de-excitations not included in the ${\widetilde{\sum}}_{i=1}^K \sigma_{\rm int}(\mathfrak{h}_i)$ term.
In both expansions, only the amplitudes corresponding to the first term on their right-hand sides are optimized in the self-consistent quantum flow procedure. 
For the flows using $(4e,4o)$-type spaces, the $\sigma_M(R)$ operator in Eqs.(\ref{expa1}) and (\ref{expa2}) (in the vein of Eq.(\ref{aux10})) is defined through  simple second-order-type estimates of triply excited cluster amplitudes, whereas 
amplitudes corresponding to the 
${\widetilde{\widetilde{\sum}}}_{i=K+1}^M \sigma_{\rm int}(\mathfrak{h}_i)$ term
are defined through first-order-type estimates of singly and doubly excited/de-excited amplitudes and second-order-type estimates for triply excited/de-excited amplitudes (a schematic representation of the corresponding diagrams are shown in Fig.\ref{fig:subf}, where vertices correspond to one- and two-body interactions of the downfolded Hamiltonian in the normal product form, and horizontal dotted lines represent resolvents constructed from diagonal elements of the corresponding Fock operator, which is defined through the dressed interactions). 
We will refer the algorithm described in Eq.(\ref{expa2}) above to as the sub-QFlow. 

The $\mathcal{S}_{\rm sub-QFlow}$ may be evaluated using various criteria that provide some form of a-priori knowledge about the importance of a particular active space. 
Promising ways to select active spaces have been discussed in Refs.
\cite{olsen1988determinant,fleig2001generalized,ivanic2003direct,sato2015time,ma2011generalized}.
Since these are the exploratory studies of sub-QFlow, we will select active spaces for sub-QFlow based on the observation of the first cycle of the genuine QFLow procedure, which includes all active spaces of a predefined form. 
%
%


\section{Results}

As a benchmark system for illustrating the performance of DUCC-QFlow and DUCC-sub-QFlow procedures, we chose an H8 linear chain of hydrogen atoms defined by a single parameter $R_{\rm H-H}$ corresponding to a distance between adjacent hydrogen atoms. In all calculations, we employed cc-pVDZ basis set \cite{dunning1989gaussian} (composed of 40 basis set functions). We performed simulations for $R_{\rm H-H}=2.0$ a.u. and $R_{\rm H-H}=3.0$ a.u. corresponding to weakly and strongly correlated ground states, respectively. The target active space is defined through 9 lowest-lying restricted Hartree-Fock (RHF) orbitals in both cases. 

\begin{figure}
\includegraphics[scale=0.35]{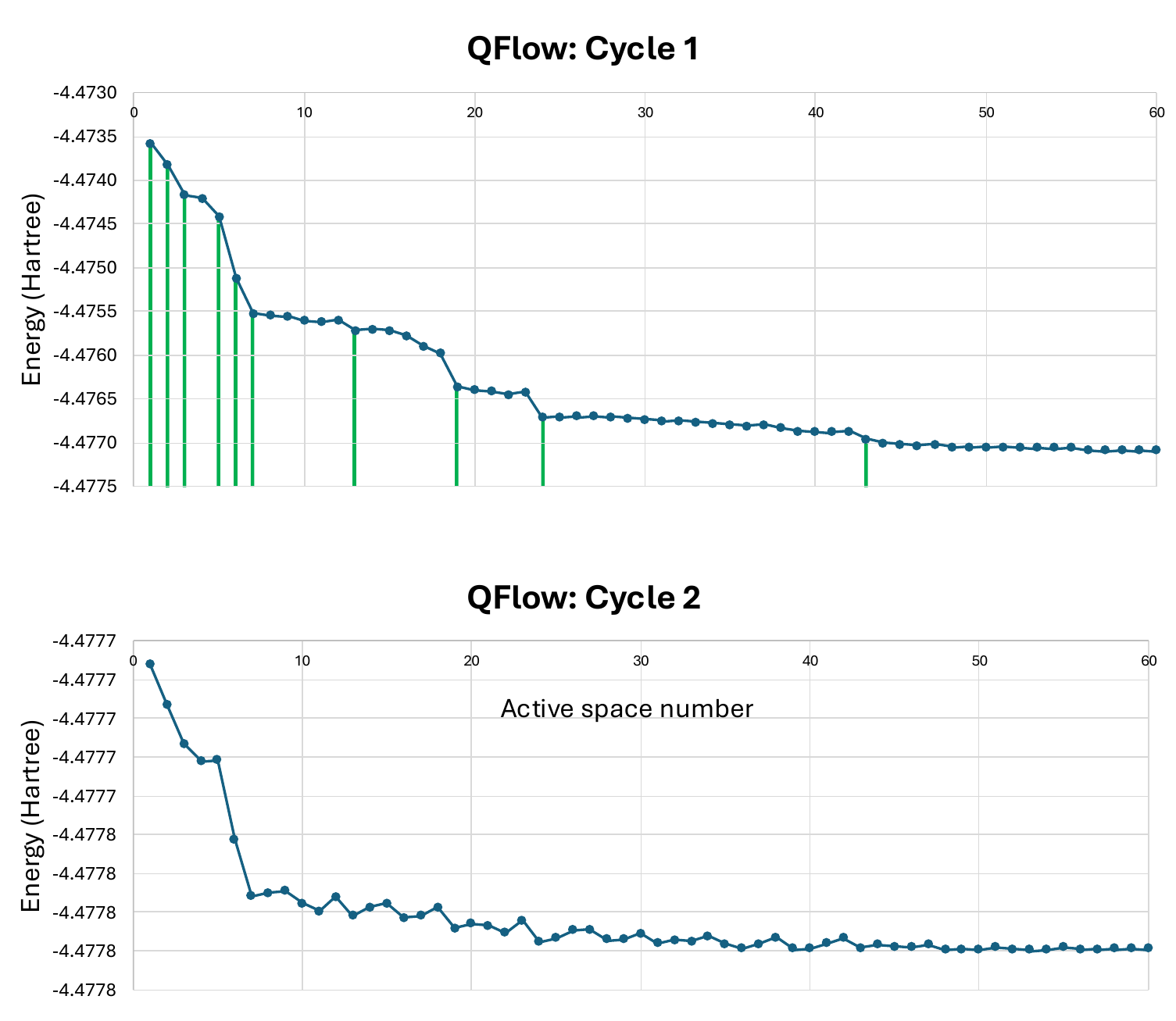}
\caption{
Energies of active space problems in the first two cycles of QFlow for H8 9-orbital model for $R_{H-H}=2.0$ a.u.
}
\label{fig:selection2au}
\end{figure}

\begin{figure}
\includegraphics[scale=0.40]{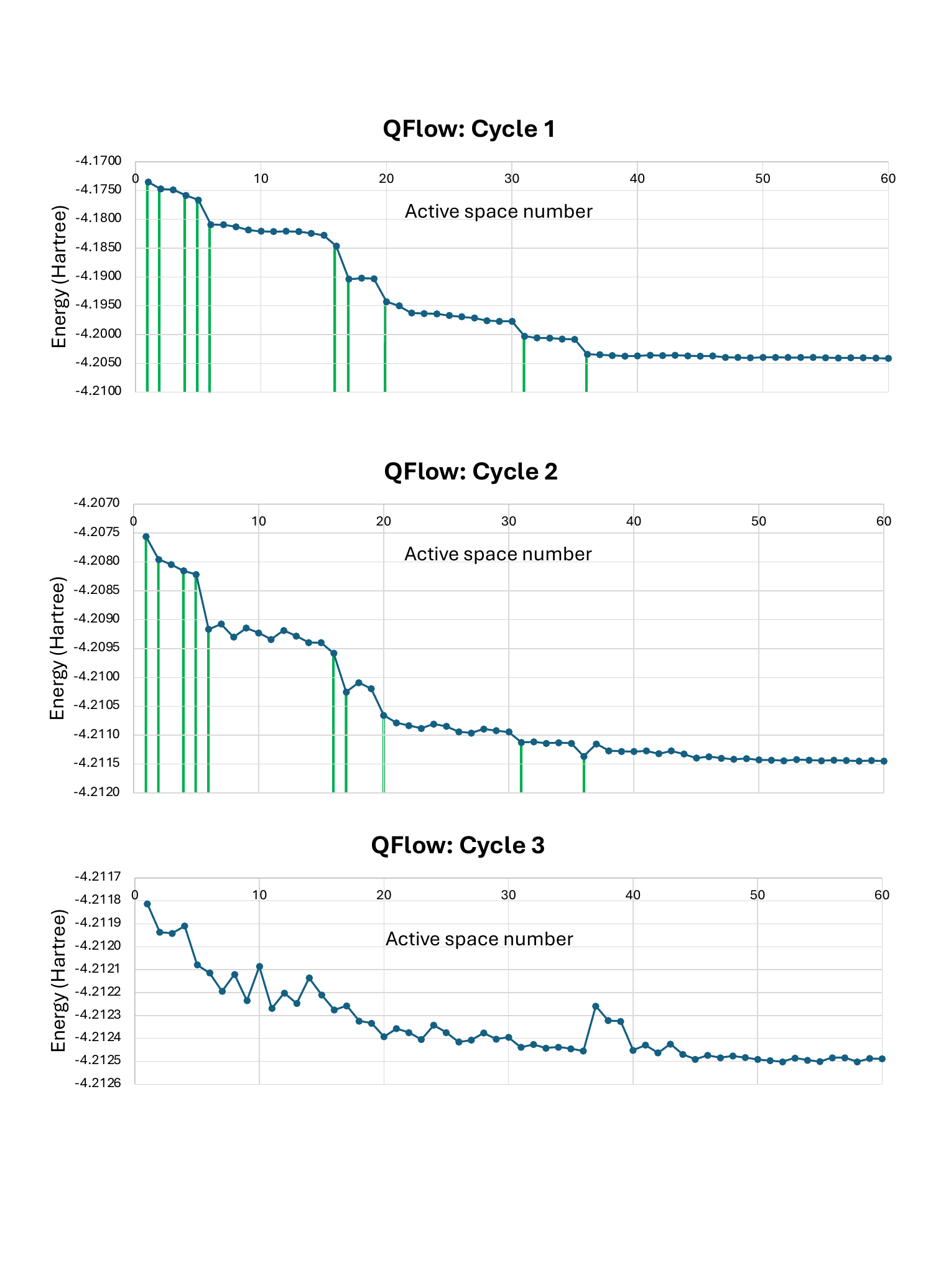}
\caption{
Energies of active space problems in the first three cycles of QFlow for H8 9-orbital model for $R_{H-H}=3.0$ a.u.
}
\label{fig:selection3au}
\end{figure}

In Figs.\ref{fig:selection2au}-\ref{fig:selection3au}, we collected energies of several cycles of the DUCC-QFlow procedure. For 2.0 a.u and 3.0 a.u. cases, one can observe a common pattern emerging. For example, in the first cycle, only several active spaces contribute to noticeable energy changes in the flow (marked by green vertical lines). In this way, we identified ten active spaces for both geometries, which are included in the self-consistent sub-QFlow procedure. Additionally, as shown in Fig for 3.0 a.u. where the convergence of the QFLow algorithm is slower, the same pattern is valid for cycles 1 and 2. 

The DUCC-QFLow and DUCC-sub-QFlow results are shown in Table \ref{tab1}. 
First, we collated ground-state energies of RHF, configuration interaction (CI), and CC methods obtained with the full inclusion of all cc-pVDZ RHF orbitals. These energies are compared to the results of exact diagonalization (ED), DUCC-QFlow, and DUCC-sub-QFLow procedures in the target active space. The VQE simulations for H8 in the cc-pVDZ basis set would require roughly 80 qubits. 
Since the DUCC-QFlow utilizes (4e,4o)-type active spaces, these simulations require eight qubits. For both geometries using the DUCC-QFlow approach, we were able to optimize 1,100 wave function parameters defining expansion (\ref{var2}), which would be  hardly achievable 
using 80 qubits of currently available hardware. 

In our DUCC-QFLow and DUCC-sub-QFlow simulations, effective Hamiltonians for active spaces are approximated based on rank-1 Trotter formula and given by Eq.(\ref{var5}) and 
the VQE optimization uses a simple gradient estimate:
\begin{equation}
\frac{\partial E(\mathfrak{h}_i)}{\partial \theta(\mathfrak{h}_i)_k} \simeq
\langle
\Psi_{\rm int}(\mathfrak{h}_i)|
[H^{\rm eff}(\mathfrak{h}_i),\tau_k(i)]
|\Psi_{\rm int}(\mathfrak{h}_i)\rangle \;,
\label{ducc6}
\end{equation}
where $\tau_k(i)$ is a corresponding combination of the strings of a creation/annihilation operators associated with the  $\theta(\mathfrak{h}_i)_k$ amplitude in the $\sigma_{\rm int}(\mathfrak{h}_i)$ operator, and 
\begin{equation}
|\Psi_{\rm int}(\mathfrak{h}_i)\rangle = e^{\sigma_{\rm int}(\mathfrak{h}_i)} |\Phi\rangle \;.
\label{var8}
\end{equation}
Instead of performing complete optimization for each active space for a given QFlow cycle, we perform only one optimization step based on a single-shot gradient evaluation given by Eq.(\ref{ducc6}).

From Table \ref{tab1}, one can notice that for strongly correlated case (3.0 a.u.),  a perturbative breakdown of the CCSD(T) formalism can be observed, where the corresponding energy is located significantly below the CCSDTQ and variational CISDTQ energies. The CCSDT approach discloses similar behavior.   The DUCC-QFlow energies agree with the CCSDTQ results with errors not exceeding 3 milliHartree for both geometries. It should be compared to the ED energies obtained through diagonalization of bare Hamiltonians in 9-orbital target space, where analogous differences amount to 122 and 83 milliHartree for 2.0 a.u. and 3.0 a.u., respectively. 
At the same time, the DUCC-sub-QFlow method reproduces the performance of the DUCC-QFlow approach for 2.0 a.u. (with an accuracy of 0.6 milliHartree). For 3.0 a.u., the DUCC-sub-QFlow and DUCC-QFlow are bigger (4.4 milliHartree), and the DUCC-sub-QFlow yields results of the CISDTQ quality. We should emphasize that the number of active spaces in the DUCC-sub-QFLow is significantly reduced with respect to the original DUCC-QFlow method. 
These approaches can be further improved by making more accurate estimates for $\sigma_M(R)$. 

%
%

\renewcommand{\tabcolsep}{0.2cm}
\begin{table}
    \centering
       \caption{Comparison of the CI and CC results for the H8 system in the cc-pVDZ basis set with ground-state energies of the 9-orbital model of DUCC-QFlow and DUCC-sub-QFlow.}
    \begin{tabular}{l c c }
    \hline \hline \\[-0.2cm]
  Method & $R_{\rm H-H}=2.0$a.u. & $R_{\rm H-H}=3.0$a.u.\\
  \hline \\[-0.2cm]
\multicolumn{3}{c}{Full-space formulations}  \\
RHF            & -4.2874  & -3.9546 \\[0.1cm]
CISD           & -4.4585  &  -4.1608 \\[0.1cm]
CISDT          & -4.4626  &  -4.1658 \\[0.1cm]
CISDTQ         & -4.4760  &  -4.2085   \\[0.1cm]
CCSD           & -4.4718  & -4.2110 \\[0.1cm]
CCSD(T)        & -4.4758  & -4.2179 \\[0.1cm]
CCSDT          & -4.4767  & -4.2203 \\[0.1cm]      
CCSDTQ         & -4.4767  & -4.2143\\[0.1cm] 
\multicolumn{3}{c}{Target active-space formulations} \\
CISD(act)      & -4.3510  &  -4.0981   \\[0.1cm]
CISDT(act)     & -4.3521  &  -4.1002   \\[0.1cm]
CISDTQ(act)    & -4.3543  &  -4.1284   \\[0.1cm]
CCSD(act)      & -4.3533  &  -4.1317   \\[0.1cm]
CCSD(T)(act)   & -4.3541  &  -4.1337   \\[0.1cm]
CCSDT(act)     & -4.3543  &  -4.1353   \\[0.1cm]
CCSDTQ(act)    & -4.3543  &  -4.1310   \\[0.1cm]
ED             & -4.3543  &  -4.1311   \\[0.2cm]
DUCC-QFlow         &  -4.4749   &  -4.2116 \\[0.1cm]
DUCC-sub-QFlow$^a$ &  -4.4743   &  -4.2072 \\[0.1cm]
%
\hline \hline
    \end{tabular}
    \label{tab1}
\end{table}

\section{Conclusion}

We demonstrated that the QFlow algorithm based on selecting the most relevant active spaces supplemented with the perturbative definition of the amplitudes not included in the sub-flow provides an adaptive alternative to QFlow algorithms based on including all fixed-type active spaces. In particular, we give an example where a 6-fold selection in the number of active spaces leads to nearly the same result for the weakly correlated variant of the H8 system and results of the CISDTQ quality for its strongly correlated variant. An essential aspect of this research is associated with a 10-fold reduction in the required qubits. It is also worth mentioning that using rather a modest number of qubits, we were able to use the DUCC-QFlow algorithm to optimize 1,100 variational parameters. 

We believe that the general concepts described in this paper can be further refined and used to efficiently capture sparsities characterizing quantum systems. Another important aspect of our research will be finding more accurate ways to characterize amplitudes not included in the self-consistent flow procedure. This is a crucial area for future exploration and could significantly enhance the effectiveness of our approach. 

\section{Acknowledgement}
This work was supported by the ``Embedding QC into Many-body Frameworks for Strongly Correlated Molecular and Materials Systems''  project, which is funded by the U.S. Department of Energy, Office of Science, Office of Basic Energy Sciences, the Division of Chemical Sciences, Geosciences, and Biosciences (under FWP 72689) and by Quantum Science Center (QSC), a National Quantum Information Science Research Center of the U.S. Department of Energy (under FWP  76213).
This work was partially supported by the U.S. Department of Energy, Office of Science, Basic Energy Sciences, Division of Materials Sciences and Engineering, Theoretical Condensed Matter Physics Program, FWP 83557.
This work used resources from the Pacific Northwest National Laboratory (PNNL).
PNNL is operated by Battelle for the U.S. Department of Energy under Contract DE-AC05-76RL01830.

\bibliography{ref}

\end{document}